# Spatial database implementation of fuzzy region connection calculus for analysing the relationship of diseases

Somayeh Davari
Department of Electrical and Computer Engineering
Isfahan University of Technology
Isfahan, Iran
somayehdvr@gmail.com

Nasser Ghadiri
Department of Electrical and Computer Engineering
Isfahan University of Technology
Isfahan, Iran
nghadiri@cc.iut.ac.ir

*Abstract*— Analyzing huge amounts of spatial data plays an important role in many emerging analysis and decision-making domains such as healthcare, urban planning, agriculture and so on. For extracting meaningful knowledge from geographical data, the relationships between spatial data objects should be analyzed. An important class of such relationships are topological relations such as connectedness or overlap between regions. While real-world geographical regions such as lakes or forests do not have exact boundaries and are fuzzy, most of the existing analysis methods neglect this inherent feature of the topological relations. In this paper, we propose a method for handling the topological relations in spatial databases based on fuzzy Region Connection Calculus (RCC). The proposed method is implemented in PostGIS spatial database and is evaluated in analyzing the relationship of diseases as an important application domain. We also used our fuzzy RCC implementation for fuzzification of the skyline operator in spatial databases. The results of the evaluation show that our method provides a more realistic view of the spatial relationships and gives more flexibility to the data analyst to extract meaningful and accurate results in comparison with the existing methods.

## I. Introduction

For many decision-making tasks, huge amounts of spatial and location-related data must be analyzed. The relationships between geographical areas are modeled using topological relationship models. The Region Connection Calculus (RCC) is a well-known and widespread model for the topological relations [1]. RCC is mainly used for reasoning about the available topological information. One of the main characteristics of this calculus that distinguishes it from the related approaches is the generality of RCC. Using an arbitrary universal set *U* of regions, the topological relations with an arbitrary reflective and symmetric relation *C* in *U* are defined as connection. The visual form of a set of RCC relations is shown in Fig. 1. In particular, note that the EC (externally connected) relation models adjacency, which can also be used to infer TPP (tangential proper part), NTPP (non-tangential proper part) and EQ (equal) relations. In different applications, geographic regions can be modeled with different methods, and the connection between them can be defined based on the specific method.

When using the RCC relations in spatial decision-making and analytics, it is usually assumed that the regions are well-defined entities, i.e. they are defined by exact boundaries. On the other hand, many geographical regions are inherently ill-defined. For example, although political areas, such as countries, states and counties, have been officially defined by exact boundaries, in many domains where people refer to them in their everyday communications (e.g., the national and local places), the regions have no exact boundaries. The basic RCC model fails to model such real-world objects, but extending the RCC with the fuzzy topological relations can be a good solution for dealing with this imprecision. Fuzzy topological relations are applicable in many fields including path tracking algorithms based on fuzzy relations [2], medical diagnosis of patient records, extracting topological relations from the web, image interpretation [3], robot control and navigation [4], brain MRI segmentation [5] and soil science [6] among many others applications domains.

There have been noticeable research efforts performed on fuzzy spatial topological relations and valuable progress has been made. Various methods for modeling fuzzy spatial features and fuzzy relations between them have been proposed and evaluated. Most definitions of the fuzzy topological relations have been developed based on either fuzzy RCC, four-intersection or nine-intersection matrix. However, applying the fuzzy RCC model to real-world spatial databases is a challenging task. Implementation of the fuzzy RCC relations in Geographic Information Systems (GIS) will help to improve the user interface compared to most of the existing GIS systems [7].

For practical applications of the fuzzy RCC relations, one of the most powerful, open source database management systems that supports spatial data is PostGIS. It can be installed on PostgreSQL DBMS and gives the user a rich set of features for handling spatial data. Many software products can use PostGIS for their database management needs. While crisp (non-fuzzy) RCC relations in some spatial databases such as PostGIS have been already implemented but fuzzy RCC relations have not been implemented yet.

In this paper, we propose a method to implement the fuzzy RCC relations in PostGIS. By implementation of the fuzzy RCC relations, they can be used in many applications. One of the important applications can be finding the spatial relations of diseases which is evaluated in this paper.

As another interesting application, the fuzzy RCC can be exploited to fuzzify the *skyline* operator. So far, approaches to making database systems more flexible to support the user needs have been proposed [8]. One of the most useful approaches is the skyline operator [9]. This operator takes a database $D$ of tuples or $n$-dimensional points and returns a set of tuples or non-dominated points in $D$. A tuple $u$ is said to dominate another tuple $u'$ if $u$ is at least as good as $u'$ in all dimensions, and superior in at least one dimension. With the spatial skyline, users can find the nearest points or regions to their intended points or regions. In this paper, the proposed fuzzy RCC method is also exploited for the fuzzy skyline problem. Real-world problems such as spatial analysis of diseases can be solved with use of this operator better than existing methods.

It should be noted that one of the most important classes of diseases is malignant diseases. Nowadays, malignant diseases are important in healthcare and require more investigation and consideration. The frequency of the malignant diseases in Iran has been increasing and many health authorities have been focused on improving the required methods. There are different reasons for increasing frequency of malignant diseases including new eating habits, increased tobacco consumption, population increase, and the increasing age of demographic structure. Although environmental pollution is one of the most important cause of certain malignant diseases, is still requires more consideration and analysis. For example, approximately one million ton of lead is added annually to the soil all over the world that great part of which is produced by atmospheric dusts, scattering of ashes, chemical fertilizers used in agriculture, industrial activities, and urban waste. As environmental factors, industrial plants also play an effective role in increasing pollution and, consequently, spreading diseases. Due to its many industrial workshops, Isfahan Province is prone to industrial pollutants. This province is also an important agricultural region where using chemical fertilizers has led to increasing levels of lead in the soil. Lead absorption from contaminated land via agricultural produce is one way this element can enter the food chain. Deposition of lead in human body causes numerous diseases and disorders including certain malignant diseases.

Through providing lead dispersion maps and observing the spatial distribution of malignant diseases in Isfahan Province using fuzzy RCC and fuzzy skyline, one can establish the relationship between lead and occurrence of malignant diseases [10].

The rest of this paper is organized as follows. Part II is an overview of the related work. Section III presents our proposed approach for implementing the fuzzy region connection calculus. In section IV of the experimental results using the real data sets are presented and compared with the non-fuzzy RCC relationships of PostGIS. Section V concludes the paper and points to some future directions.

II. RELATED WORK

Our proposed fuzzy RCC method is to be exploited for the spatial analysis of an important real-world problem in healthcare. Essential elements of descriptive epidemiology include three factors: "person", "place", and "time". However, in the past decades, epidemiology has been mainly focused on time and person, largely ignoring the role of place [11]. Providing various maps from geographical patterns of diseases has been devised and practiced for more than a hundred years. Development and improvement of geographical systems in the past 40 years have led to more advanced possibilities for studying geographical patterns [12]. This has in turn led to increased debates on the application of geographical information systems in studies related to public health and epidemiology.

Application of geographical information systems is important due to the fact that epidemiology studies have shifted their focus from the group to the individual. Also, the need for suitable technologies to study epidemiologic data has increased as well as the application of these systems. In some studies, an overall view of the health problems can be provided via geographic information systems (GIS) [13].

*A. Using GIS for epidemiological and public health studies*

GIS applications in human health studies can be divided into two groups. First, those epidemiologic applications which mostly involve the study of geographical changes to find the extent the disease has spread, as well as the possible causes of such change [14]. These systems can also be used for monitoring and controlling various diseases, and can be used for monitoring diseases transmitted by carriers [15]. The second group are the application related to designing health-care systems, particularly those which can be readily accessed [14]. GIS can also be used for finding the most suitable place for establishing health care centers such as hospitals, and evaluating some projects like population-based treatment of tuberculosis [13].

*B. The Frameworks used for Investigating the Relationship between Diseases*

Although RCC provides an attractive framework for modeling topologic relations, it has limited applications in real-world situations where spatial features are affected by vagueness. In one study, a generalization of RCC is proposed which provides the possibility of defining spatial relations between the vague regions [16]. For this purpose, the spatial relations were modeled as fuzzy relations. To support such a spatial argument based on these relations, certain important properties as well as a transmissibility table were presented. It also shows how to model vague spatial information when vague regions are represented as fuzzy sets. Upon completion of this study, standard RCC models were introduced by Schockaert et al. [17]. In another study, an RCC generalization was introduced based on the fuzzy set theory which indicates how argumentation tasks such as satisfying and checking the implication could overcome the difficulties encountered in linear programming [18].

Non-fuzzy RCC relations have been already implemented in certain spatial databases. However, fuzzy RCC relations have not been implemented in PostGIS yet. In this paper, the fuzzy relations in the PostGIS spatial databases were implemented. By implementing these relations in the spatial PostGIS data bases, we can exploit them in different applications such as the spatial skyline query.

To make the skyline more flexible, fuzzy skylines are introduced [19,20]. Different types of fuzzy skyline were examined in [21], as well as their characteristics for dealing with uncertainty. For the disease analysis field, making skyline more flexible and using larger scales are the two applicable motives chosen from among the five suggested motives. By making skyline more flexible, we can obtain skylines large enough to meet the requirements of the analyst. Thus, the analyst can investigate the tuples obtained from executing the spatial queries and conduct the required research. Larger scales can help to diversify the tuples obtained from skyline and make the skyline more comprehensible.

### III. THE PROPOSED METHOD

#### A. Region Connection Calculus

The RCC model is introduced in [22]. RCC considers a primary relation $C(x,y)$ to be correct when the $x$ and $y$ have a common point. Based on this primary connect relation, other RCC relations are defined between two regions.

The RCC-8 model consists of eight relations: DC, EQ, PO, EC, PO, EC, TPP, TPPi, NTPP, NTPPi where TPP and NTPP are inverses of the TPP and NTPP relations respectively. Fig. 1 illustrates the RCC-8 relations. This collection includes all the details and shows a set of relationships between two regions [22]. RCC has been proposed as a widepsread method for spatial reasoning in GIS [22].

#### B. Fuzzy Region Connection Calculus

Defining topological relations can start with a fuzzy relation $C$ in an universe $U$ of regions. For regions $u$ and $v$ in $U$, $C(u,v)$ is a degree between [0,1] that $u$ and $v$ are connected to each other with it.

In comparison with the non-fuzzy RCC, $C$ is a fuzzy reflective and symmetric relation. For generalization of a fuzzy connection, a t-norm is used as a mapping in $[0,1]^2$-$[0,1]$ having the symmetric, associative and increasing properties. The boundary condition $T(a,1) = a$ for all $a$ in $[0,1]$ is also established. It can be shown that such a mapping behaves as a connection operator directly. In the following Łukasiewicz t-norm being used:

$$T_W(a, b) = \max(0, a+b-1) \quad (1)$$

Other popular t-norms are the minimum $T_M$ and product $T_P$

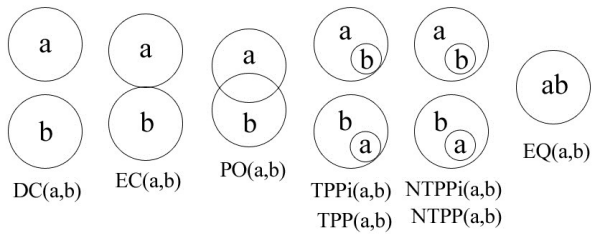

Fig. 1. The topological relations : RCC-8 [1]

defined by $T_M(a,b) = \min(a,b)$ and $T_P(a,b) = a \cdot b$ for all $a$ and $b$ in $[0,1]$.

There are several methods to obtain the connection of fuzzy spatial features. A commentary that is especially noticeable is based on nearness. Consider fuzzy relation $R(\alpha,\beta)$ in $\mathbb{R}^n$ defined for each p and q in $\mathbb{R}^n$ ($\alpha,\beta \geq 0$).

$$R_{(\alpha,\beta)}(p,q) = \begin{cases} 1, & \text{if } d(p,q) \leq \alpha \\ 0 & \text{if } d(p,q) > \alpha + \beta \\ \frac{\alpha+\beta-d(p,q)}{\beta} & \text{otherwise } (\beta \neq 0) \end{cases} \quad (2)$$

Where d is the Euclidean distance in $\mathbb{R}^n$. For two points $p$ and $q$ in $\mathbb{R}^n$, by considering appropriate values for the parameters $\alpha$ and $\beta$, $R_{(\alpha,\beta)}(p,q)$ shows the degree to which $p$ is near to $q$. Note that if $d(p,q) \leq \alpha$ then $p$ and $q$ are considered to be quite near, while if $d(p,q) > \alpha + \beta$ then $p$ and $q$ are considered not near. There are linear gradual transition between these values; for example, if $d(p,q) = \alpha + \frac{\beta}{2}$ then $p$ and $q$ are close to 0.5 degree. Using this notion of closeness of the points, a connection can be interpreted based on the closeness between vague regions. In particular, consider $A$ and $B$ as two normalised fuzzy sets in $\mathbb{R}^n$ with a bounded support. The grade $C_{(\alpha,\beta)}(A,B)$ indicates amount of connection between $A$ to $B$ is defined as follows:

$$C_{(\alpha,\beta)}(A,B) = \sup_{p \in \mathbb{R}^n} T(A(p), \sup_{p \in \mathbb{R}^n} T(R_{(\alpha,\beta)}(p,q), B(q))) \quad (3)$$

This shows the amount of closeness between some points of $A$ to some points of $B$. According to the above definition of connection and by considering regions as fuzzy regions, fuzzy RCC relations are definable by connection [18].

#### C. The Proposed method for implementing fuzzy RCC relations

The process of implementing the fuzzy RCC relations is shown in Fig. 2. It requires implementing the fuzzy connect relation first. After that, using the fuzzy connection, other fuzzy RCC relations can be implemented. For implementing the fuzzy connection, regions should be defined as fuzzy areas so the connection between these fuzzy regions can be evaluated.

The following code fragment performs this operation:

```
CREATE TYPE FuzzyRegion AS (
core geometry ,
supportRadius float );
```

It defines a new type in PostGIS that for each fuzzy region associates every geometry with a number. The geometry is the core of fuzzy region and the number indicates the support radius of the fuzzy region, as can be seen in Fig. 3.

Thereby, the intended fuzzy region can be accessed and used in relations. For obtaining the degree of membership of each point in this fuzzy region, if the point is in the *Support Radius* area, the following formula can be applied:

$$\text{Membership} = 1 - \frac{\text{Distance(Point,Core)}}{\text{Support Radius}} \quad (4)$$

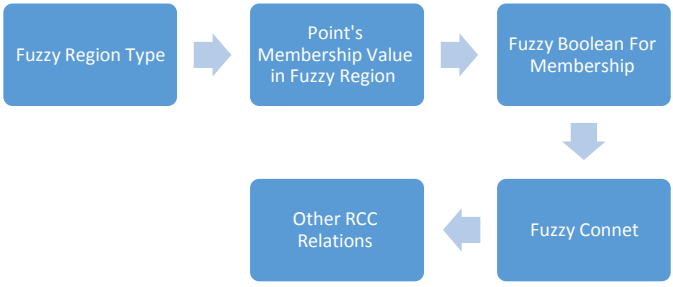

Fig. 2. The process of implementing fuzzy RCC relations

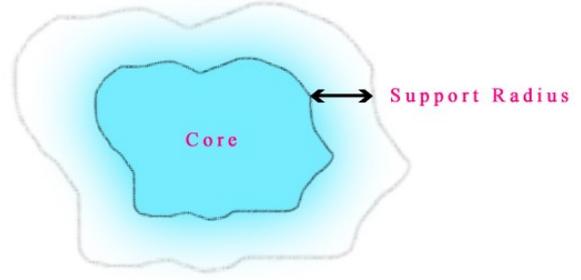

Fig. 3. A fuzzy region

Otherwise, if the point is outside of the zero area and if it is in the core, a value of one is assigned as the membership value of the point in the fuzzy region. In Algorithm 1, a function is defined that receives this fuzzy region and a point, and returns the point's membership value in the fuzzy region based on Eq. (4).

**Algorithm 1.** *FuzzyMembership in pgSQL*

```
1.  create function Fmembership
    (a FuzzyRegion, b geometry)
2.  returns fuzzy_boolean as
3.  $body$
4.  declare
5.  Membership float;
6.  begin
7.   if (a.supportRadius = 0) then
8.          if ST_DWithin(a.core, b, 0) then
9.                  return 1;
10.         else return 0;
11.         end if;
12.  else
13.         Membership := 1 - (ST_distance ( b ,
     a.core ) / 1. supportRadius );
14.         if (Membership < 0) then
15.                 return 0;
16.         else return Membership;
17.         end if;
18.  end if;
19.  end;
20.  $body$
21.  language 'plpgsql';
```

This function receives a fuzzy region and a geometry of a point. If the point is farther from the fuzzy region's area, it returns zero as the membership value of the point in the fuzzy region, otherwise it returns a number as the point's membership in the fuzzy region based on its distance from the geometry of the fuzzy region's center. This number is a fuzzy membership floating point value between zero and one and is defined as:

**CREATE DOMAIN** fuzzy_boolean
AS FLOAT CHECK (VALUE BETWEEN 0 AND 1);

These functions are used in the implementation of the fuzzy connection. Algorithm 2 shows the proposed method for implementing the fuzzy connection.

This function receives two regions with *Support Radius SR* and *DD* as the number of divisions in each dimension and $\alpha$ and $\beta$ related to fuzzy connection relation initially, checking if the region is not empty. Then it defines regions as fuzzy regions. The number of divisions is a flexible parameter and the user can increase it for more accuracy

**Algorithm 2.** *FuzzyConnect*

```
Input: G1, G2, α, β, DD, SR
Output: FuzzyBoolean
1.  if G1 is empty or G2 is empty then return 0 else
2.    divide G1 into DD*DD parts
3.    divide G2 into DD*DD parts
4.    for each part of G1 loop
5.        for each part of G2 loop
6.            compute distance between them
7.            if (D <= α) then R ← 1
8.            else if (D > α+β) then  R ← 0
9.            else R ← ((α+β-D)/β)
10.           p ← (XMin of the G1's part, YMin
    of the G1's part)
11.           q ← (XMin of the G2's part, YMin
    of the G2's part)
12.           AArray ← T(R,FMembership(G2,q))
13.       end loop
14.       AMax ← max(AArray)
15.       BArray ← T(FMembership(G1,p),AMax)
16.  end loop
17.  end if
18.  return max(BArray)
```

of the results. Lines 4 and 5 are divisions of the regions, and then fuzzy connection relations are implemented. By using the fuzzy connection, other fuzzy RCC relations can also implemented easily.

*D. Evaluation methods*

To evaluate the proposed method, the fuzzy connection relation with distance function that is available in the PostGIS can be used. By assigning random numbers to the regions as *Support Radius*, the connectivity can achieved and compared with the distances.

The rest of the relations are based on fuzzy connectivity and have the same properties. The evaluation of the other RCC relations can be considered in a future work.

IV. EVALUATION

*A. Experiment Design*

For evaluating the proposed method, we implemented it in PostGIS. The results are obtained from the implementation of the method on a system with a 2 GHz Core i7 CPU and 6 GB of RAM. The datasets we used were the lead distribution dataset in Isfahan province, as used in [10]. We recreated and cleaned the data using the ArcGIS software. These data are geometric. The second dataset was Isfahan urban areas dataset which includes 31 Isfahan province urban areas. This collection was downloaded from the NaturalEarth website (http://www.naturalearthdata.com/).

*B. Implementation of method on diseases and lead datasets*

The functions implemented in the previous section for fuzzy RCC on the lead distribution and Isfahan urban areas datasets were executed and evaluated. As shown in TABLE I, the overlap of the Isfahan cities with the lead was obtained. By comparing this rate and the amount of diseases for each city, it can be observed that the results are closely related and similar but more details in comparison with the results of [10] are obtained. The study revealed that the distribution of lead in the city of Isfahan is directly related to the amount of diseases but they did not obtain the *amount* of this association. The implementation of fuzzy RCC on the data obtained and the amount is confirmed this association. This study uses the overlap relation and the degree of overlap between any city and distribution of lead was computed. It can show that, for example, the distribution of lead in the Isfahan is more than other cities and the amount of disease in this city is more than the other areas in this province.

TABLE I. COMPARISON OF DISEASES, CITY AND OVERLAP

| City | Overlap | Skin | Breast |
|---|---|---|---|
| *Isfahan* | 0.7375 | 1 | 1 |
| *Lower Semirom / Dehaghan* | 0.5890 | 0.066 | 0.055 |
| *Mobarakeh* | 0.3254 | 0.052 | 0.036 |
| *Lanjan* | 0.2604 | 0.004 | 0 |
| *Falavarjan* | 0.1743 | 0.012 | 0.003 |
| *Khansar* | 0.0616 | 0.029 | 0.008 |
| *Ardestan* | 0.0597 | 0.091 | 0.067 |
| *Natanz* | 0.0393 | 0.079 | 0.059 |
| *Borkhar and Meymeh* | 0.0240 | 0.075 | 0.052 |
| *Golpayegan* | 0.0112 | 0.031 | 0.013 |
| *K Homeynishahr* | 0.0108 | 0.375 | 0.145 |
| *Kashan* | 0.0070 | 0.127 | 0.101 |
| *Chadegan* | 0.0032 | 0.049 | 0.026 |
| *Semirom* | 0.0014 | 0.001 | 0 |
| *Fereydunshahr* | 0 | 0.026 | 0.008 |
| *Nayin* | 0 | 0.105 | 0.068 |

*C. Finding the nearest areas*

Fig. 4 shows the results of applying the proposed method for fuzzy connection and distance function on the Isfahan urban areas dataset. In this implementation, distances and fuzzy connections of each region with the region with gid=124 has been studied. The number eight is considered for number of divisions and α and β are equal to zero and 0.01 respectively. Support Radius is the fuzzy limited area for each region and is filled with a random number between zero and three. The horizontal axis represents areas and the vertical axis is the connectivity and distance.

Fig. 5 shows the results of applying five steps of the fuzzy connectivity on the Isfahan urban areas dataset. In this implementation, by assigning different minimums for connectivity, those regions satisfying the minimums have been added.

At first stage the blue region with gid=124 satisfied the minimum connectivity requirement. At the next stage the yellow and the blue regions satisfied the minimum connectivity requirement. At the third phase, the red areas are added to these areas. At the forth stage, the green regions satisfied the requirement and finally, the pink one has been added to the other regions.

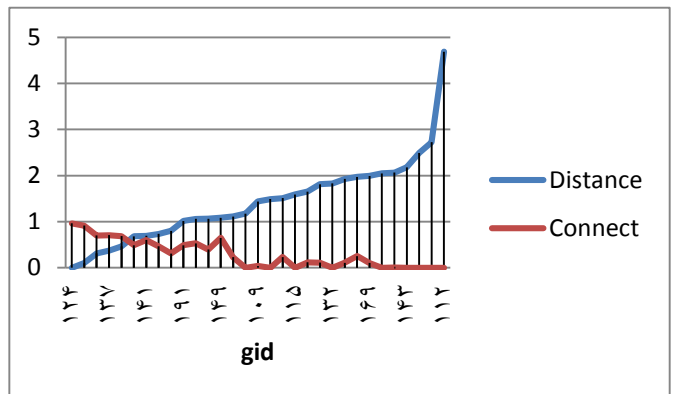

Fig. 4. Comparison of fuzzy connectivity and distances in data. The *gid* in this diagram is the number identifying each area.

*D. Implementation of the proposed skyline on the Isfahan urban areas dataset*

Suppose that we want to investigate the impact of industrial plants on people and on the weather of the urban areas, For instance the areas that are closer to oil refinery and Isfahan steel company. An example can be analyzing amount of each type of diseases in these areas and to investigate the association between the diseases and its proximity to refineries. To find these areas, the basic skyline operator was applied on the Isfahan urban areas dataset which returns only a single area that was closer to the oil refinery and to the Isfahan steel company. To improve this, and by applying the proposed fuzzy skyline to this dataset, we can obtain an arbitrary larger amount of areas to be used in the study.

The TABLE II shows the results of implementation of the proposed method. In all results of this table except the two last rows, a support radius of one is being considered. This amount is 0.5 and 0.01, respectively, for the two last rows. In the rows with bold letters just the *R* value has been applied but *R* and *C* are applied in the next rows. All runtime numbers in this table are in millisecond. As can be seen in the TABLE II, any arbirtrary large amount of areas for the study can be achieved by the proposed method. However, as it can be seen, the fuzzy skyline running time is higher than the basic skyline.

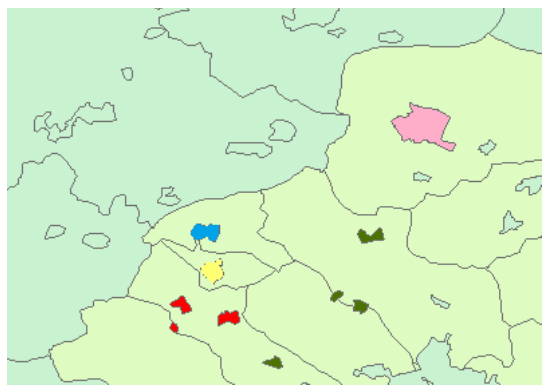

Fig. 5. Runing the fuzzy connection with five different thresholds

*E. Evaluation and interpretation of results*

By observing and comparing the results of the fuzzy connectivity function and distance function, and by considering the fuzzy boundaries of the regions, the accuracy of the proposed function is clear. The proposed algorithm has also a reasonable running time. In addition, in its implementation, by specifying a lower number of divisions, the results can be achieved the less running time. Although the lower number of division slightly reduces algorithm's accuracy but can highly increase the execution speed.

TABLE II. COMPARISON OF SKYLINE RESULTS

| $T_{Fuzzy}$(ms) | $T_{Sum}$(ms) | $N_{Fuzzy}$ | $N_{Sum}$ | α | β | $Min_C$ |
|---|---|---|---|---|---|---|
| **141** | **141** | **2** | **2** | **0** | **0.3** | **0.01** |
| **171** | **171** | **4** | **4** | **0** | **0.5** | **0.01** |
| **181** | **181** | **6** | **6** | **0** | **0.8** | **0.01** |
| **191** | **191** | **9** | **9** | **0** | **1** | **0.01** |
| **195** | **195** | **6** | **6** | **0** | **1** | **0.2** |
| **285** | **285** | **30** | **30** | **2** | **2** | **0.01** |
| 16901 | 16901 | 4 | 4 | 0 | 0.3 | 0.01 |
| 19882 | 19882 | 10 | 10 | 0 | 0.5 | 0.01 |
| 21926 | 21926 | 15 | 15 | 0 | 0.8 | 0.01 |
| 23480 | 23480 | 17 | 17 | 0 | 1 | 0.01 |
| 21864 | 21864 | 14 | 14 | 0 | 1 | 0.2 |
| 25197 | 25197 | 20 | 20 | 0 | 1.5 | 0.2 |
| 28284 | 28284 | 28 | 28 | 0.5 | 1.5 | 0.2 |
| 31345 | 31345 | 30 | 30 | 1 | 1.5 | 0.2 |
| 28615 | 28615 | 28 | 28 | 1 | 1 | 0.2 |
| 26217 | 26217 | 23 | 23 | 1 | 0.5 | 0.2 |
| 22390 | 22390 | 16 | 16 | 1 | 0 | 0.2 |
| 28458 | 28458 | 26 | 26 | 1 | 1 | 0.2 |
| 16702 | 16702 | 4 | 4 | 1 | 1 | 0.2 |

## V. CONCLUSION

In this paper, to exploit the potential of fuzzy spatial region relationships, we proposed a method for implementing the fuzzy RCC in PostGIS to tackle real-world spatial analysis tasks. With fuzzy RCC we can calculate fuzzy spatial relations between fuzzy spatial features. It has been shown that fuzzy RCC has a great potential for many targets. Afterward, we evaluated this method and used it on a dataset in comparison with distance function. We also used it to study about the spatial distribution of diseases and the spatial aspects of relationships. Fuzzy RCC helped in finding the amounts of the relations that cannot be shown with non-fuzzy methods. As another application of the fuzzy RCC it was used for fuzzification of the skyline operator based on user interests. The benefits of this model are flexibility, user-friendliness, resonable speed and usable results. With this implementation we can also find the skyline for fuzzy regions. This fuzzy skyline model was used in a healthcare and diseases research problem. The proposed methods provides many benefits, yet having the possibility for further improvements such as performance improvement and appying in various domains other than healthcare.